\newcommand{\PaperTitle} {Lost in the Prefix: Revisiting IP Geolocation Accuracy Across Networks and Geographies}

\documentclass[10pt,twocolumn,letterpaper]{article}

\usepackage[margin=0.75in]{geometry}
\usepackage{epsfig, endnotes, graphicx, xcolor, xspace, subcaption}
\usepackage[all]{nowidow} %

\usepackage{multirow}  %
\usepackage{booktabs}  %
\usepackage{tabularx}  %
\usepackage{array}     %
\usepackage{arydshln}  %
\usepackage{hyperref}
\usepackage{newtxtext,newtxmath}

\newcommand\paragraphb[1]{\noindent{\bf{#1}}}

\begin{document}

\title{\PaperTitle}

\author{
  Syed Tauhidun Nabi \\
  Virginia Tech \\
  \texttt{tauhidun@vt.edu}
  \and
  Jocelyn Bliton \\
  Virginia Tech \\
  \texttt{jocelynb23@vt.edu}
  \and
  Tijay Chung \\
  Virginia Tech \\
  \texttt{tijay@vt.edu}
  \and
  Shaddi Hasan \\
  Virginia Tech \\
  \texttt{shaddi@vt.edu}
}
\date{}  %

\maketitle

\begin{abstract}
IP geolocation databases are widely used in research, policy, and industry, yet their accuracy across network types and geographies remains poorly characterized. We present a large scale evaluation of four major providers (MaxMind GeoLite2, IPinfo, IP2Location, and DB-IP) using ground truth from RIPE Atlas and UNICEF Giga across 175 countries. We find that mobile networks exhibit median errors more than 10 times higher than fixed networks across all providers (179--207~km vs.\ 3--16~km), and that Global South regions show significantly higher failure rates than Global North: Asia exceeds 53--61\% and Africa 66--72\%, compared to 9--20\% in Europe. We trace both gaps to a shared structural source: provider prefixes in mobile networks and Global South geographies are more likely to be coarser than BGP announcements, and approximately 70\% of mobile prefixes span more than 100~km geographically. Our findings point to prefix granularity as a common explanatory factor: coarser prefixes consistently produce the highest errors regardless of provider, network type, or geography.

\end{abstract}

\section{Introduction}
\label{sec:intro}

IP geolocation databases map IP addresses to physical locations and are widely used across research, industry, and policy. Measurement platforms, researchers, and policy makers use them to associate performance data with geographic communities and identify underserved populations~\cite{mlab, USBroadbandUsage2022, nabi2024red, saxon2022gps}. Content delivery networks use them for traffic steering and latency optimization~\cite{poese2011ip}, and advertising platforms use them for geographic targeting~\cite{favia2022advertising}. Despite this pervasive use, the accuracy of these databases outside the conditions under which they have historically been evaluated remains poorly understood.

Prior evaluations~\cite{poese2011ip, saxon2022gps, gharaibeh2017look} have documented meaningful errors in IP geolocation, but their ground truth is drawn predominantly from fixed broadband networks in North America and Europe. This leaves two significant blind spots. First, mobile networks, where subscribers' IP addresses are assigned by carrier-grade NAT infrastructure \cite{imc2016richter} that may aggregate traffic across large geographic regions, have received little systematic attention. Second, the Global South, where mobile connectivity is often the \emph{primary} access medium and where IP address allocation practices may differ substantially from wealthier regions, is nearly absent from existing benchmarks. Whether current databases are accurate enough to support the research, policy, and operational decisions that depend on them in these contexts is an open question.

We address this gap with a large scale measurement study of four major IP geolocation providers: MaxMind GeoLite2~\cite{maxmind_geolite2}, IPinfo~\cite{ipinfo}, IP2Location~\cite{ip2location}, and DB-IP~\cite{dbip}. We evaluate accuracy against ground truth drawn from two complementary sources. RIPE Atlas~\cite{staff2015ripe} provides our primary instrument, offering stable, well-characterized vantage points that are dense in North America and Europe but span multiple continents, predominantly on fixed networks. UNICEF Giga school connectivity measurements \cite{giga2023, gigameter2023} augment our dataset with coverage of the Global South and mobile networks that RIPE Atlas alone cannot provide at scale; to our knowledge, this is the first use of the Giga dataset as a ground truth instrument for IP geolocation evaluation, introducing it as a new resource for the network measurement community. Together, our ground truth comprises 37,302 observations spanning 175 countries collected over a one-month window, with explicit coverage of mobile, fixed, and satellite access network types.

We address two research questions:
\begin{itemize}
    \item \textbf{RQ1:} How does IP geolocation accuracy vary across access network types?
    \item \textbf{RQ2:} How does IP geolocation accuracy vary across geographies at global scale?
\end{itemize}

Our analysis reveals consistent and large accuracy gaps along both dimensions. Mobile networks exhibit median errors an order of magnitude higher than fixed networks, with all four providers failing at similar rates on mobile, pointing to a common rather than provider-specific cause. Global South regions show significantly higher failure rates than Global North across all providers and all continents. We trace both gaps to a shared structural source: provider prefixes in mobile networks and Global South geographies are substantially coarser than BGP announcements, and approximately 70\% of mobile prefixes span more than 100~km geographically.

This paper makes the following contributions. First, we present the first large scale evaluation of IP geolocation accuracy across both fixed and mobile networks and Global South geographies, using 37,302 georeferenced observations spanning 175 countries. Second, we introduce the UNICEF Giga~\cite{giga2023} school connectivity measurements as a new ground truth resource for evaluating geolocation accuracy, providing coverage of mobile networks and the Global South absent from prior benchmarks. Finally, we show that BGP prefix granularity is a consistent explanatory factor for the Global South accuracy gap, and a contributing factor for the mobile accuracy gap, supported by within-prefix geographic spread analysis and BGP prefix size comparisons across all geolocation providers.

\section{Datasets and Methodology}
\label{s:datasets}

\subsection{Ground Truth Dataset}
\label{s:datasets:truth}
We construct ground truth from two complementary sources.

\paragraphb{RIPE Atlas}~\cite{staff2015ripe} provides our primary instrument, comprising 10,294 IPv4 and 5,716 IPv6 probe records collected on March 23, 2026, filtered to probes with operator-reported coordinates, stable IP addresses over the preceding 30 days, and excluding IPs appearing at multiple probes to remove potential NAT or VPN addresses (323 rows removed, 16,010 retained across 10,561 unique probes). RIPE Atlas probes are dense in North America and Europe, and only approximately 1\% carry a mobile network tag, limiting mobile network coverage in our ground truth. We note that a small fraction (1.5\%) of RIPE Atlas probes misreport their geolocation, with violations disproportionately concentrated in low-coverage regions such as southern Africa~\cite{izhikevich2024trust}; we treat probe-reported coordinates as ground truth with this caveat.

\paragraphb{UNICEF Giga} school connectivity measurements augment our dataset with Global South and mobile network coverage that RIPE Atlas cannot provide at scale~\cite{giga2023, gigameter2023}. Giga runs NDT7 speed tests~\cite{mlab} from GPS-reported school locations across 33 countries. We apply four sequential filters: (1) a one-month measurement window (February 15 to March 15, 2026); (2) removal of measurements where school GPS coordinates are missing; (3) deduplication to unique (IP, school) pairs, 
and (4) exclusion of IPs appearing at more than one school within the one-month measurement window, a conservative heuristic to identify NAT/CGNAT addresses where the same IP serves multiple geographic locations. The filtered Giga dataset comprises 21,292 unique (IP, school) pairs across 4,872 schools in 27 countries. Combined with RIPE Atlas, our ground truth totals 37,302 observations spanning 175 countries, of which 74.7\% are IPv4 and 25.3\% are IPv6.

\paragraphb{IP Reuse and Unit of Analysis.} We treat individual observations as our unit of analysis, where each observation is defined as a unique (IP address, vantage point) pair. For RIPE Atlas, each observation corresponds to a single connected probe with a stable IPv4 or IPv6 address and operator-reported coordinates. For Giga, each observation corresponds to one unique (IP, school) pair within the measurement window. IP reuse in Giga takes two forms: a single school observed with multiple IPs over time, consistent with DHCP reassignment at a fixed location; and a single IP observed at multiple schools, consistent with NAT, CGNAT, or VPN infrastructure aggregating traffic from geographically dispersed locations~\cite{imc2016richter, cloudflare2025cgnat}. The former preserves valid ground truth, as each IP corresponds to that school's location at the time of measurement. The latter introduces ground truth uncertainty, as the IP termination point may be far from any individual school. We conservatively exclude IPs appearing at more than one school within the one-month measurement window to remove likely NAT, CGNAT, and VPN addresses.

\subsection{Geolocation Provider Databases}
Our evaluation covers data from four geolocation providers: MaxMind GeoLite2~\cite{maxmind_geolite2}, IPinfo~\cite{ipinfo}, IP2Location~\cite{ip2location} DB11, and DB-IP~\cite{dbip} Lite.\footnote{MaxMind GeoLite2 and DB-IP Lite are freely available. We obtained a research license to IPinfo's commercial product. IP2Location provided access to their commercial DB11 product after we contacted them with preliminary results using their free LITE product (\S\ref{s:discussion}).}
All four snapshots were collected in the fourth week of March 2026 (March 25 for MaxMind, IPinfo, and IP2Location; March 27 for DB-IP), closely following our ground truth collection window (February 15 to March 23, 2026), to minimize risk of drift.
Each entry in a geolocation database consists of contiguous IP address ranges, with one predicted location assigned per range, rather than per IP address. 
We refer to these ranges as \textit{geolocation prefixes}.

\subsection{Mobile Vantage Point Classification}
Although user tags allow us to classify RIPE Atlas vantage points as mobile vs fixed, the Giga dataset does not directly label the access network type of each measurement. 
We thus derive the network type from the client ASN recorded by the M-Lab platform for each NDT speed test run from these vantage points. 
First, we map each ASN to its organization name using the CAIDA AS2Org dataset~\cite{caida_as2org}.
Second, an AI agent classifies each organization by performing a web search and inspecting (in priority order) whether the organization has been assigned an MCC/MNC code~\cite{itu-t-e212} (a unique identifier for mobile network operators), the services and plans outlined on the organization's website, and news articles about the organization's service offerings. 
Based on this information, the agent classifies each operator by access technology type (Mobile only, Fixed only, Both, or Satellite), a dominance label (mobile dominant, fixed dominant, or balanced) where applicable, a confidence score, and a reasoning statement supporting its classification.
Finally, a domain expert on the author team manually validates and corrects all agent classifications before they are applied to the dataset.

For analyses comparing mobile and fixed network accuracy, we define mobile observations as those classified as Mobile only, or Both with mobile dominant traffic. Fixed observations are those classified as Fixed only, or Both with fixed dominant traffic. Observations classified as Both with balanced dominance cannot be attributed to a single access type at the observation level and are excluded from mobile versus fixed comparisons but retained in geography-level analyses. Satellite observations are included in overall accuracy analysis but excluded from mobile-versus-fixed comparisons. This classification yields 5,358 mobile, 18,310 fixed, 12,071 Both, and 1,196 satellite observations.

\subsection{BGP Prefix Data and Classification}
\label{s:datasets:bgp}
To compare geolocation prefix granularity against globally announced routing prefixes, we use the CAIDA RouteViews Prefix-to-AS dataset~\cite{caida_pfx2as, routeviews}
snapshot dated March 15, 2026, containing 1,082,764 IPv4 prefix entries. 
For each observed IP, we compare its geolocation prefix size to the corresponding BGP-announced prefix, classifying the relationship into either \textbf{Exact} matches, \textbf{Smaller (finer)} when the geolocation prefix is smaller than the BGP-announced prefix, \textbf{Larger (coarser)}, or \textbf{No Match} if there is no overlap with an announced prefix.
We examine how the distribution of these categories varies in \S\ref{s:prefix}.

\section{Results}
\label{s:results}

\subsection{Accuracy by Network Type}
\label{s:results:network}

To characterize how geolocation accuracy varies across access network types, we compare error distributions for fixed and mobile observations across all four providers.

\begin{figure}[h!]
  \centering
      \includegraphics[width=0.9\linewidth]{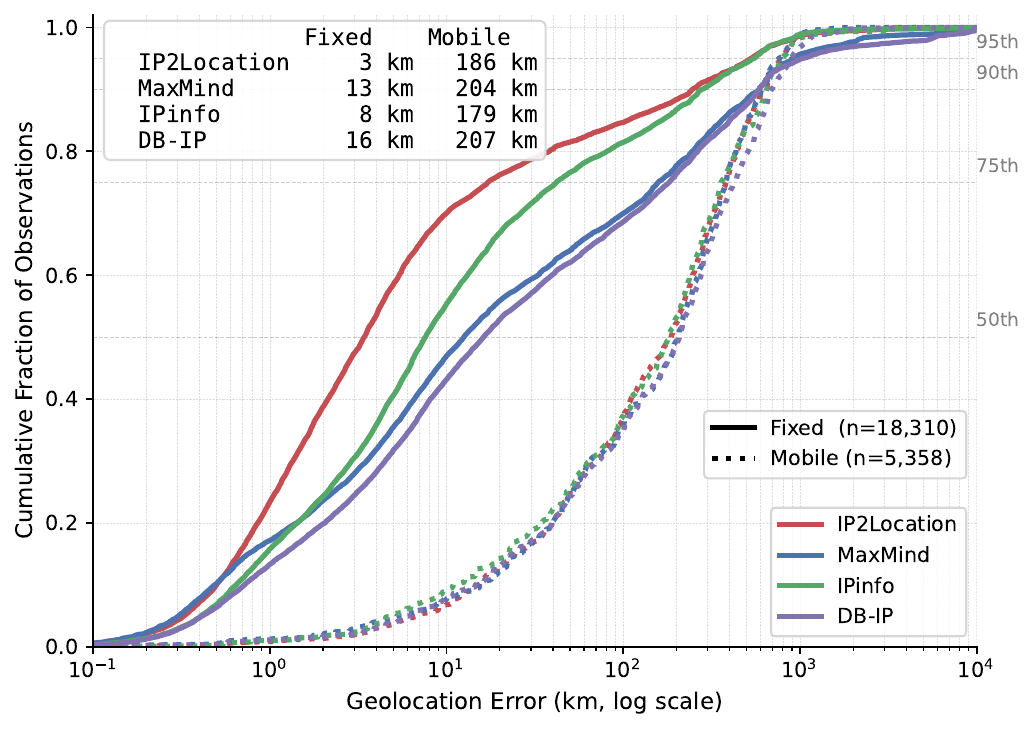}
  \caption{Geolocation error for fixed and mobile networks. Mobile networks show median errors more than 10 times higher than fixed across all providers.}
  \label{fig:cdf_error_fixed_vs_mobile}
\end{figure}

Figure~\ref{fig:cdf_error_fixed_vs_mobile} shows the CDF of geolocation error for fixed and mobile observations across all four providers. Fixed networks achieve median errors of 3--16~km across providers (IP2Location 3~km, IPinfo 8~km, DB-IP 16~km, MaxMind 13~km), though p90 errors exceed hundreds of kilometers even for fixed networks, indicating that a long tail of poorly geolocated addresses exists regardless of network type. Mobile networks exhibit dramatically higher errors: median errors range from 179 to 207~km across providers (IPinfo 179~km, IP2Location 186~km, DB-IP 207~km, MaxMind 204~km), an order of magnitude higher than fixed network performance.

Critically, all four providers fail at similar rates on mobile networks, with their CDF curves converging in the mobile regime despite diverging substantially on fixed networks. This convergence suggests the mobile accuracy gap is not attributable to differences in provider methodology but reflects a common property of mobile network addressing. We investigate this source in \S\ref{s:prefix}.

\subsection{Accuracy by Geography}
\label{s:results:geo}

To examine how geolocation accuracy varies across geographies, we compute the failure rate per continent, defined as the fraction of observations where geolocation error exceeds 100~km --- corresponding roughly to the scale at which errors become consequential for regional analyses.

\begin{figure*}[t]
  \centering
  \includegraphics[width=0.8\linewidth]{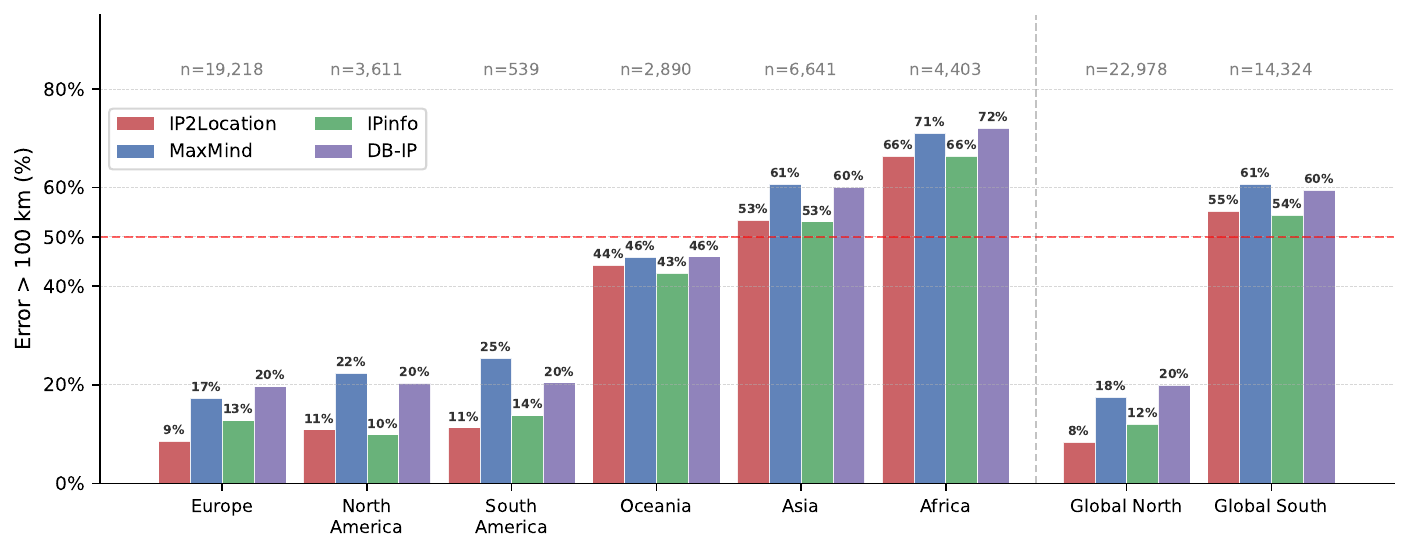}
\caption{Geolocation failure rate (error $>$ 100~km) by continent and global region across four providers. Asia and Africa exceed 50\% across all providers; Europe and the Americas stay below 25\%.}
  \label{fig:failure_rate}
\end{figure*}

Figure~\ref{fig:failure_rate} shows the geolocation failure rate by continent and global region across all four providers. A clear and consistent geographic disparity emerges across both dimensions, and this pattern holds across all four providers.

Asia and Africa are the worst affected regions. Across all providers, Asia failure rates range from 53 to 61\% and Africa from 66 to 72\%. Oceania shows intermediate failure rates of 43 to 46\%, while the Americas (8 to 22\%) and Europe (9 to 20\%) remain well below the 50\% threshold. The consistency of these rankings across all four providers indicates the geographic disparity is not provider-specific. At the global region level, Global South observations show failure rates 2 to 3 times higher than Global North across all providers (GS: 54--61\%, GN: 8--20\%).

\paragraphb{Statistical significance.} To test whether the Global North versus Global South accuracy gap is statistically significant, we compute country-level median errors and apply the Mann-Whitney U test~\cite{mann1947test}, with rank-biserial correlation~\cite{kerby2014simple} as the effect size measure, excluding countries with fewer than 20 observations, leaving 44 Global North and 44 Global South countries. 
Table~\ref{tab:stat_test} shows the results. All four providers show a statistically significant gap, with effect sizes ranging from large (IP2Location, r=0.666) to small (DB-IP, r=0.256), indicating that while the geographic gap is universal, its magnitude varies across providers.

\begin{table}[h!]
\small
\centering
\begin{tabular}{lccc}
\toprule
\textbf{Provider} & \textbf{MW p-value} & \textbf{Effect size (r)} \\
\midrule
IP2Location & $<$0.001$^{*}$ & 0.666 \\
IPinfo      & $<$0.001$^{*}$ & 0.429 \\
MaxMind     &    0.005$^{*}$ & 0.322 \\
DB-IP       &    0.019$^{*}$ & 0.256 \\
\bottomrule
\multicolumn{3}{l}{\footnotesize $^{*}$significant at $\alpha = 0.05$} \\
\end{tabular}
\caption{Global North vs.\ Global South country-level median error (Mann-Whitney U test,
$\alpha = 0.05$). Effect size $r$ is the rank-biserial correlation.}
\label{tab:stat_test}
\end{table}

\paragraphb{Country-level misassignment.}
Beyond distance errors, wrong-country assignments are rare across all four providers (fewer than 1\% of observations per provider), but the cases that do occur tend to be concentrated in specific prefixes rather than distributed randomly. Europe and the Americas account for the largest absolute counts, driven primarily by cross-border spillover among geographically proximate countries (e.g., Germany assigned to France). In contrast, Asia shows the fewest misassignments despite having the highest failure rates, suggesting Asian errors manifest as large within-country distance errors rather than cross-border placement. Notable cases include Fiji observations assigned to the United States by IP2Location and MaxMind (167 observations each, all satellite), and Turks and Caicos observations assigned to Jamaica by IP2Location and DB-IP (15 observations each). Full wrong-country assignment counts by continent are shown in Appendix (Figure~\ref{fig:country_mismatch}).
\section{What Limits Accuracy?}
\label{s:prefix}
Our results in \S\ref{s:results} illustrate that all four geolocation providers face similar accuracy gaps in both mobile networks and in the Global South, despite each using independent geolocation methodologies.
To investigate this further, we evaluated the \emph{geolocation prefix} structure of each database across these dimensions, specifically the \emph{spread} among observed vantage points within a geolocation prefix as well as the relationship between geolocation prefix size and announced BGP prefixes.

\subsection{Within-Prefix Geographic Spread}
\label{s:prefix:spread}

\begin{figure}[h!]
  \centering
  \includegraphics[width=0.9\linewidth]{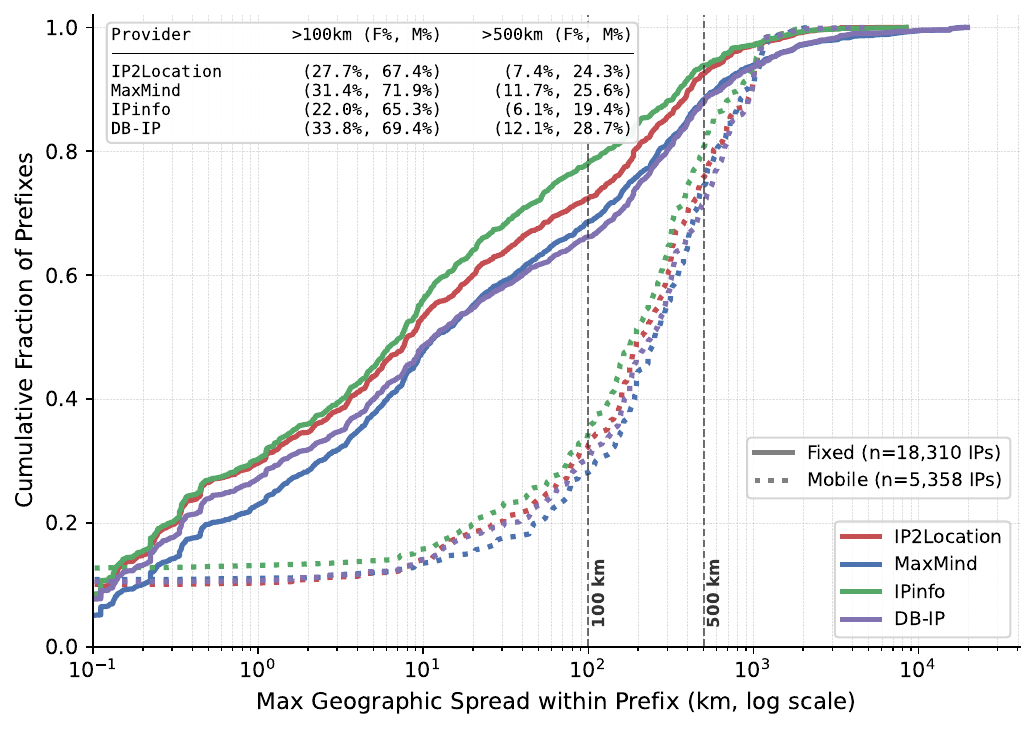}
  \caption{Within-prefix geographic spread for fixed and mobile networks. Geolocation prefixes covering mobile networks tend to span larger areas than those from fixed networks.}
  \label{fig:cdf_spread_fixed_vs_mobile}
\end{figure}

We compute the spread of each geolocation prefix directly: for every prefix with at least two unique ground truth IPs, we calculate the maximum pairwise Haversine distance among all unique vantage point coordinates assigned to that prefix.
A geolocation prefix with a large spread is inaccurate since it groups IP addresses together that are actually located across a broad area.
Across all geolocation providers, a substantial fraction of geolocation prefixes have large spreads: between 32 and 40\% of all geolocation prefixes span more than 100~km, and 9 to 14\% span more than 500~km.
Within-prefix spread is significantly larger in mobile networks compared to fixed (Figure~\ref{fig:cdf_spread_fixed_vs_mobile}): approximately 70\% of mobile prefixes span more than 100~km.
Within-prefix spread thus drives part of the mobile accuracy gap observed in \S\ref{s:results:network}; although this is expected given the nature of mobile networks and prior work~\cite{saxon2022gps}, our work is the first to quantify this behavior.
The same pattern holds geographically. Table~\ref{tab:prefix_spread_geo} compares the distribution of wide prefixes (spread $>$ 100~km) between the Global North and Global South. 
Wide prefixes are roughly twice as prevalent in the Global South, similarly driving the accuracy gap observed in \S\ref{s:results:geo}.
IP addresses that fall within geolocation prefixes with large spread will suffer poor accuracy due to coarse database granularity; this disproportionately impacts addresses on mobile networks and in the Global South.

\subsection{Geolocation vs.\ BGP Prefix Granularity}
\label{s:prefix:bgp}
We next compare each geolocation prefix against the most specific BGP-announced prefix covering its IP address range. 
Our objective is to determine whether providers operate at a coarser, finer, or equal granularity compared to routing infrastructure, and whether this pattern varies across network types and geographies in ways that explain the accuracy gaps observed in \S\ref{s:results}.

\begin{table}[t!]
\small
\centering
\begin{tabular}{lcccc}
\toprule
 & \multicolumn{2}{c}{\textbf{Global North}} &
   \multicolumn{2}{c}{\textbf{Global South}} \\
\cmidrule(lr){2-3} \cmidrule(lr){4-5}
\textbf{Provider} & \textbf{\% Wide} & \textbf{Median} &
                    \textbf{\% Wide} & \textbf{Median} \\
\midrule
IP2Location & 19.5\% & 186~km & 52.5\% & 343~km \\
MaxMind     & 30.3\% & 269~km & 55.8\% & 413~km \\
IPinfo      & 16.3\% & 196~km & 49.8\% & 315~km \\
DB-IP       & 21.8\% & 246~km & 57.9\% & 363~km \\
\bottomrule
\end{tabular}
\caption{Within-prefix geographic spread by region. Prefixes with $>$100~km spread are 2-3x more prevalent in the Global South.}
\label{tab:prefix_spread_geo}
\end{table}

\begin{table*}[t]
\small
\centering

\begin{tabular}{lrccccccccccc}
\toprule
\multirow{2}{*}{\textbf{Provider}} &
\multirow{2}{*}{\textbf{Total}} &
\multicolumn{3}{c}{\textbf{Prefix Class (\%), p50 Error (km)}} &
\multicolumn{2}{c}{\textbf{Fixed (Larger)}} &
\multicolumn{2}{c}{\textbf{Mobile (Larger)}} &
\multicolumn{2}{c}{\textbf{GN (Larger)}} &
\multicolumn{2}{c}{\textbf{GS (Larger)}} \\
\cmidrule(lr){3-5} \cmidrule(lr){6-7}
\cmidrule(lr){8-9} \cmidrule(lr){10-11}
\cmidrule(lr){12-13}
& &
\textbf{Exact} & \textbf{Smaller} & \textbf{Larger} &
\textbf{\%} & \textbf{p50 Err.} &
\textbf{\%} & \textbf{p50 Err.} &
\textbf{\%} & \textbf{p50 Err.} &
\textbf{\%} & \textbf{p50 Err.} \\
\midrule
IP2Location & 12,185
  & 24\%, 44 & 69\%, 16 &  7\%,  54
  &  8\% &   6~km &  4\% & 208~km
  &  5\% &   4~km & 13\% & 119~km \\
MaxMind     & 10,520
  & 26\%, 29 & 54\%, 30 & 19\%,  85
  & 18\% &  28~km & 29\% & 217~km
  & 13\% &  34~km & 39\% & 179~km \\
IPinfo      & 12,222
  & 19\%, 34 & 70\%, 21 & 10\%,  61
  & 10\% &   7~km & 12\% & 225~km
  &  7\% &  10~km & 18\% & 176~km \\
DB-IP       & 12,432
  & 23\%, 60 & 71\%, 41 &  6\%,  21
  &  7\% &   6~km &  3\% & 338~km
  &  5\% &   6~km & 10\% & 101~km \\
\bottomrule
\end{tabular}
\caption{BGP prefix classification, median geolocation error by class, and Larger prefix share by network type and region for all IPv4 provider prefixes. Larger (Coarser) prefixes produce the highest errors in three of four providers and are 2 to 3 times more prevalent in Global South (GS) than Global North (GN).}
\label{tab:bgp_combined}
\end{table*}

Table~\ref{tab:bgp_combined} summarizes the BGP prefix classification results, median errors by class, and the geographic breakdown of Larger prefixes across all four providers for IPv4.
Among provider prefixes observed in our ground truth, the majority are finer than their corresponding BGP announcements, indicating that providers subdivide prefixes beyond what is advertised in routing tables. 
However, a meaningful fraction aggregate across BGP boundaries (Larger): MaxMind shows the highest Larger share at 18.7\%, more than three times that of DB-IP (5.9\%). 

The median error columns reveal a consistent ordering in three of four providers: Larger prefixes produce the highest errors and Smaller prefixes produce the lowest, with Smaller median errors of 16--41~km. DB-IP is an exception where this ordering does not hold, likely reflecting its small Larger prefix pool. Full error distributions by BGP class are shown in Figure~\ref{fig:bgp_error_by_class} in the Appendix.

The rightmost columns of Table~\ref{tab:bgp_combined} show how Larger prefix share and associated errors differ between Global North and Global South. The Global South consistently shows 2 to 3 times more Larger prefixes than the Global North: MaxMind's Global South share reaches 39\% compared to 12.9\% in the North, and even DB-IP shows a two-fold difference (9.8\% vs.\ 4.6\%). The error consequences are substantial: Global South p50 errors for Larger prefixes range from 101 to 179~km across providers, compared to 4 to 34~km in the Global North. This concentration of coarse prefixes in the Global South directly explains the geographic accuracy gap observed in \S\ref{s:results:geo}, and is consistent with the within-prefix spread findings in \S\ref{s:prefix:spread}: Global South IP address space is disproportionately covered by prefixes that aggregate across BGP boundaries, likely reflecting structural differences in IP address allocation density between regions~\cite{cloudflare2025cgnat}. The Fixed and Mobile columns of Table~\ref{tab:bgp_combined} further show that mobile p50 errors within the Larger class remain substantially higher than fixed across all providers (208--338~km vs.\ 6--28~km), despite variable Larger prefix share between the two network types. This indicates that BGP prefix coarseness alone does not fully account for the mobile accuracy gap; CGNAT-driven geographic dispersion of subscribers within coarse prefixes is an additional contributing factor, consistent with the within-prefix spread findings in \S\ref{s:prefix:spread}.
\section{Discussion}
\label{s:discussion}
\paragraphb{Geolocation provider engagement.}
We contacted all four geolocation providers to discuss our results; only IPinfo and IP2Location responded to us.
IPinfo acknowledged the challenges of deploying measurement vantage points in the Global South compared to the Global North.
For prefix sizing, they noted that hijack-defense announcements can produce smaller announced BGP prefixes that do not correspond to finer geographic granularity, and that geolocation providers sometimes collapse same-location prefixes together to reduce physical database size. 
IP2Location provided access to their commercial DB11 database for comparison against the LITE version used in our initial evaluation. 
In line with prior work~\cite{saxon2022gps}, commercial version shows only modest improvement in overall median error (19 to 17~km) and p90 error (379 to 368~km), with no improvement on mobile (186~km median for both) compared to the LITE version.

\paragraphb{MaxMind accuracy radius.}
MaxMind's GeoLite2 database uniquely provides a stated accuracy radius field intended to convey confidence in each location estimate. We find that over half of observations (51\%) exceed MaxMind's stated accuracy radius, with a p90 ratio of actual error to stated radius of 10$\times$. Researchers and practitioners who use the accuracy radius field to filter or weight geolocation estimates should be aware that it substantially underestimates true error, particularly in mobile and Global South contexts.

\paragraphb{Limitations.}
RIPE Atlas probe coordinates are operator-reported with a small misreport rate concentrated in low-coverage regions~\cite{izhikevich2024trust};
Giga GPS coordinates are sourced from government stakeholders (typically Ministries of Education) through national school censuses or dedicated location collection efforts, and undergo basic validity checks by the Giga team (boundary containment, duplicate detection) but are not independently validated in most countries~\cite{giga2023}. 
These coordinates may not precisely reflect the network attachment point of each measurement, introducing a small but unquantifiable source of ground truth uncertainty. 
A large fraction of our combined ground truth (57\% of Giga observations) falls in ASNs we categorize as providing Both fixed and mobile service and is excluded from mobile versus fixed comparisons, which may limit the representativeness of our RQ1 findings. 
Our analysis covers a single one-month snapshot in early 2026; we do not assess temporal stability of geolocation prefix assignments or accuracy trends. Finally, our BGP prefix analysis is limited to IPv4.

\section{Related Work}
\label{s:related}

\paragraphb{IP geolocation methods.}
Early work on IP geolocation explored complementary approaches. Padmanabhan and Subramanian~\cite{padmanabhan2001investigation} compared database-driven methods against active measurement techniques, concluding that database methods held the greatest practical promise. Gueye et al.~\cite{gueye2004constraint} proposed constraint-based geolocation using latency measurements from distributed landmarks, Katz-Bassett et al.~\cite{katz2006towards} further refined latency-based geolocation using topology measurements from traceroutes. Despite continued refinement of active methods, commercial databases remain the dominant approach in practice due to their low cost and ease of deployment, motivating the need for accurate characterization of their limitations.

\paragraphb{IP geolocation database evaluation.}
Poese et al.~\cite{poese2011ip} conducted an early evaluation of commercial IP geolocation databases, finding median errors between tens and hundreds of kilometers.
Shavitt and Zilberman~\cite{shavitt2011geolocation} evaluated multiple databases using point-of-presence based ground truth derived from delay measurements. 
Gharaibeh et al.~\cite{gharaibeh2017look} examined router geolocation, using CAIDA Ark as ground truth. 
Saxon and Feamster~\cite{saxon2022gps} leveraged dense GPS-located vantage points in three US cities to show that IP geolocation performs well on fixed broadband but poorly on mobile networks.
Our work extends these evaluations to 175 countries with explicit mobile and Global South coverage, and provides a structural explanation for the accuracy gaps through BGP prefix analysis.

\paragraphb{Mobile network addressing and CGNAT.}
Richter et al.~\cite{imc2016richter} showed that CGNAT is near ubiquitous in mobile networks, causing many subscribers to share a single public IP address across large geographic areas. Giotsas and Fayed~\cite{cloudflare2025cgnat} further characterized CGNAT concentration in the Global South, finding that Africa and Asia have the highest user-to-IP ratios. These structural properties of mobile addressing align with the prefix coarseness we observe in \S\ref{s:prefix}.

\section{Conclusion}
\label{s:conclusion}

We presented a large scale evaluation of four major IP geolocation providers across 175 countries, using GPS-reported ground truth from RIPE Atlas and UNICEF Giga with explicit coverage of mobile networks and Global South geographies. Our analysis reveals that mobile networks exhibit median errors over 10 times higher than fixed networks across all providers, and that Global South regions show failure rates 2 to 3 times higher than Global North --- gaps that are consistent across providers and statistically significant.

Through within-prefix geographic spread analysis and BGP prefix comparison, we trace both gaps to a shared structural source: provider prefixes in mobile networks and Global South geographies are substantially coarser than BGP announcements, and a single prefix frequently covers geographic areas spanning hundreds of kilometers. These findings suggest that improving geolocation accuracy in mobile and Global South contexts requires finer prefix granularity at the provider level --- a challenge rooted in IP address allocation structure rather than database methodology alone.

Our findings have direct implications for researchers who use IP geolocation in measurement studies. 
Elevated mobile and Global South geolocation error is common to all geolocation providers we evaluate, both free and commercial, so studies cannot escape it by switching geolocation providers.
We recommend that researchers (i) avoid sub-national geographic claims for IPs whose access network or country falls in these classes, and (ii) treat the geolocation prefix size as a per-observation accuracy bound — large prefixes flag estimates that should be down-weighted or excluded.

\bibliographystyle{plain}
\bibliography{refs}

\appendix
\section*{Appendix}

\section{Ethics}

Our use of data in this paper raises no ethical concerns. All datasets used in this paper are publicly available: RIPE Atlas probe metadata is publicly accessible via the RIPE Atlas API, UNICEF Giga school connectivity measurements are available through the Giga project, and all four IP geolocation databases are available for download under free or academic licenses. No human subjects were involved in this research.

\section{Terminology}
\paragraphb{Global North and Global South.}
We classify countries into Global North and Global South following the UNCTAD country classification scheme~\cite{unctad_classification}, with country-to-region mapping following the UN M49 standard~\cite{unm49}. Global North encompasses countries in Northern America, Europe, Russia, Australia, and New Zealand. Global South encompasses countries in Africa, Asia (excluding Japan, South Korea, Singapore, and Israel, which follow UNCTAD developed economy classification), Latin America and the Caribbean, and Pacific Island nations. Our statistical comparison in Section~\ref{s:results:geo} includes 44 Global North and 44 Global South countries after excluding countries with fewer than 20 observations.

\paragraphb{Geolocation Prefix vs.\ BGP Prefix.}
A \textit{geolocation prefix} is a contiguous range of IP addresses to which a geolocation database assigns a single predicted location. Provider prefixes are defined by each database independently based on their proprietary data sources and methodology, and may not align with routing boundaries. A \textit{BGP prefix} is a block of IP addresses announced by an autonomous system via the Border Gateway Protocol, reflecting how the global routing infrastructure divides and delegates address space. The two are distinct: a provider's geolocation prefix may be finer, coarser, or exactly aligned with the BGP prefix covering the same address range. We compare these in Section~\ref{s:prefix:bgp} to assess whether providers operate at a granularity consistent with routing infrastructure.

\section{Distribution of Ground Truth Vantage Points}

\begin{figure*}[t]
  \centering
  \includegraphics[width=0.8\linewidth]{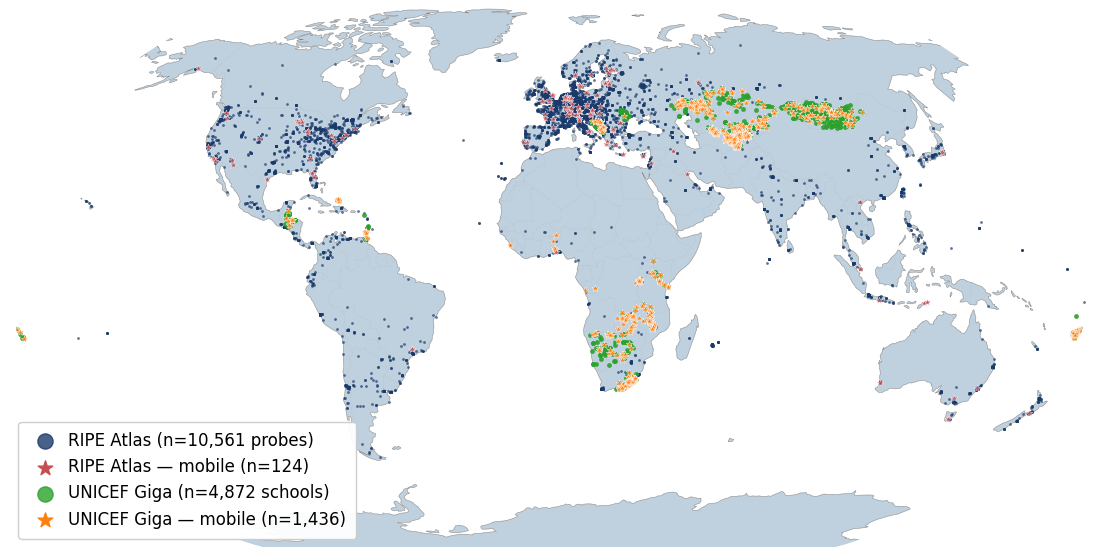}
  \caption{Ground truth vantage points from RIPE Atlas (blue circles, n=10,561 probes) and UNICEF Giga (green circles, n=4,872 schools), with mobile-tagged points shown as stars. RIPE Atlas is dense in Europe and North America but includes only 124 mobile vantage points, while UNICEF Giga provides broader Global South coverage (e.g., Central Asia, Africa, Fiji) and contributes 1,436 mobile vantage points, forming the primary source of mobile ground truth.}
  \label{fig:ground_truth_vps}
\end{figure*}

Figure~\ref{fig:ground_truth_vps} shows the geographic distribution of all ground truth vantage points used in our evaluation. The complementary coverage of RIPE Atlas and UNICEF Giga highlights a key limitation of existing measurement infrastructure: RIPE Atlas probes are concentrated in Europe and North America, while Giga extends coverage to underrepresented regions including Central Asia, sub-Saharan Africa, and Oceania. The distribution of mobile vantage points further motivates RQ1, as mobile coverage is almost entirely driven by Giga, which serves as the primary source of mobile ground truth in our dataset.

\section{Additional Results: Country-Level Misassignments}
\label{app:mismatch}

\begin{figure*}[t]
  \centering
  \includegraphics[width=0.9\linewidth]{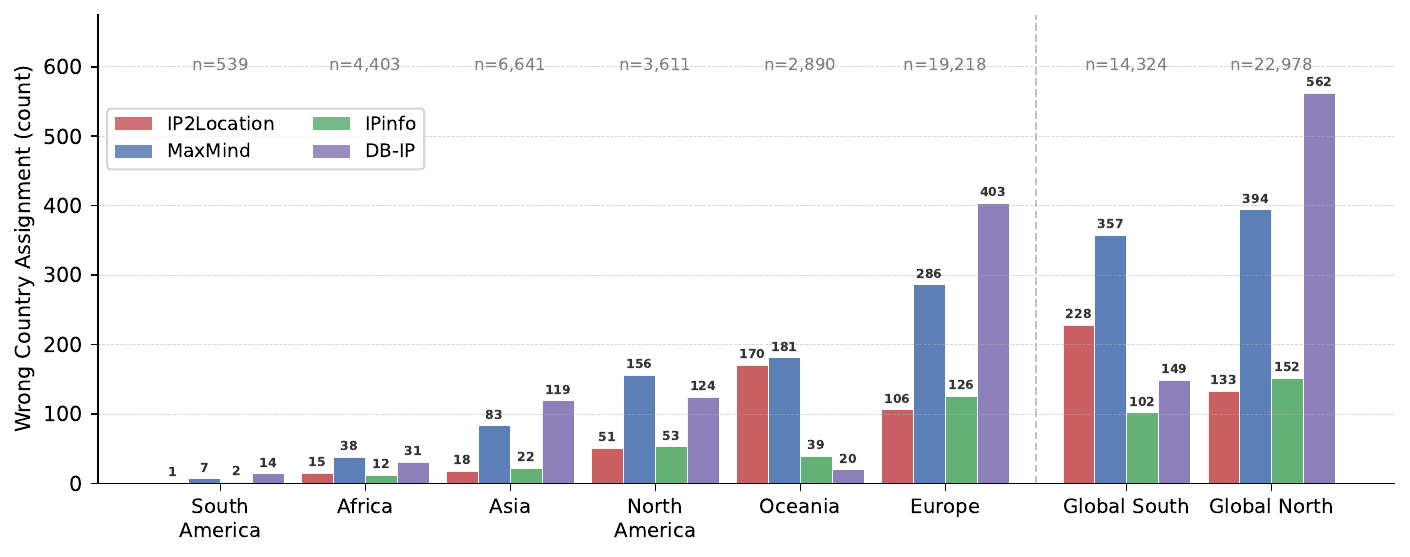}
  \caption{Wrong-country assignment count by continent across all four providers. While rare overall (fewer than 1\% of observations per provider), misassignments concentrate in Europe, the Americas, and Oceania rather than in Asia and Africa where failure rates are highest.}
  \label{fig:country_mismatch}
\end{figure*}

Figure~\ref{fig:country_mismatch} shows the full distribution of country-level misassignment counts by continent across all four providers, complementing the summary presented in Section~\ref{s:results:geo}.

\section{Geolocation Error by BGP Prefix Classification}

\begin{figure*}[t]
  \centering
  \begin{subfigure}{0.48\textwidth}
    \includegraphics[width=\linewidth]{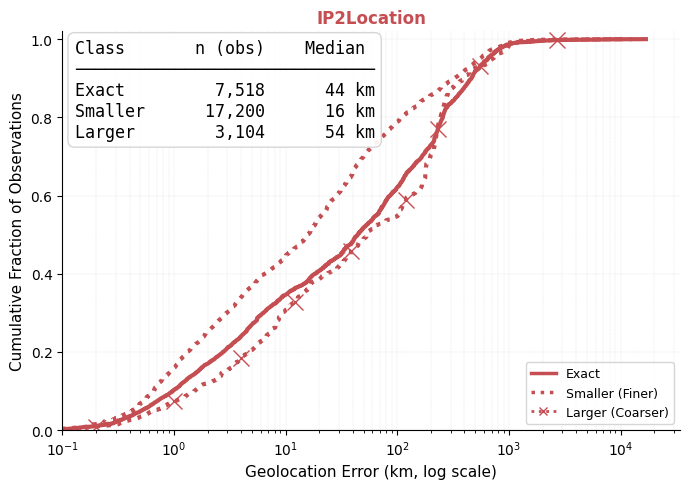}
  \end{subfigure}
  \hfill
  \begin{subfigure}{0.48\textwidth}
    \includegraphics[width=\linewidth]{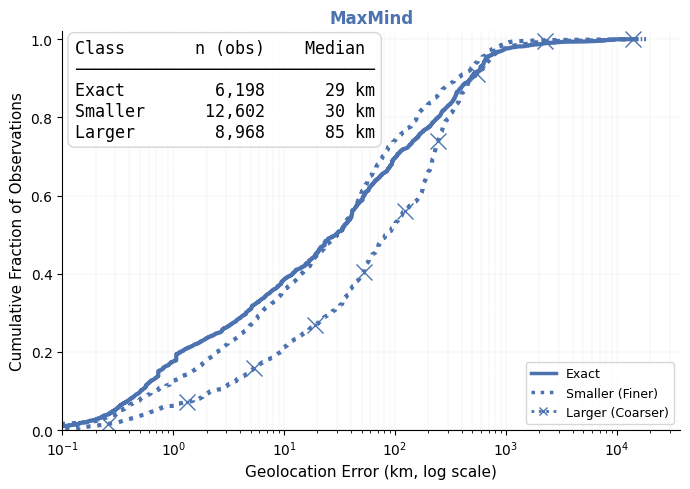}
  \end{subfigure}
  \vspace{0.5em}
  \begin{subfigure}{0.48\textwidth}
    \includegraphics[width=\linewidth]{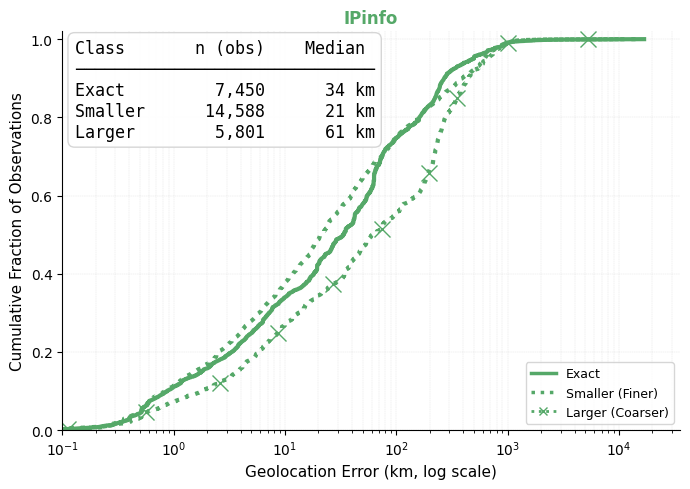}
  \end{subfigure}
  \hfill
  \begin{subfigure}{0.48\textwidth}
    \includegraphics[width=\linewidth]{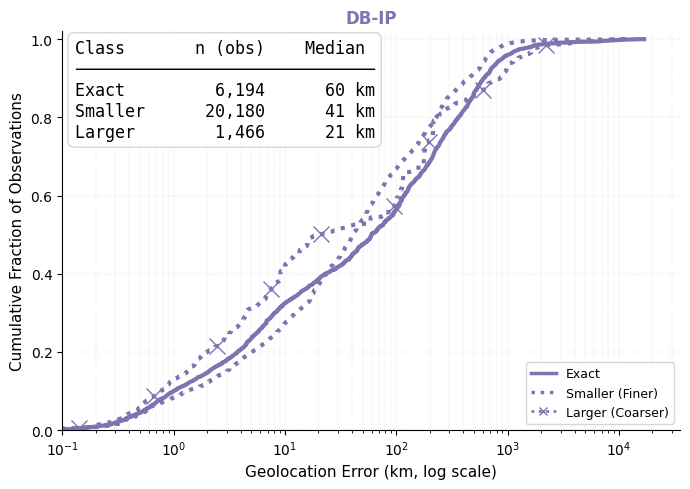}
  \end{subfigure}
  \caption{CDF of geolocation error by BGP prefix classification across all four providers (IPv4). Larger (Coarser) prefixes produce the highest median errors and Smaller (Finer) prefixes produce the lowest across three of four providers. DB-IP is an exception where this ordering does not hold, likely reflecting its small Larger prefix pool (n=1,466 observations).}
  \label{fig:bgp_error_by_class}
\end{figure*}

Figure~\ref{fig:bgp_error_by_class} shows the full error distributions stratified by BGP prefix classification for all four providers, complementing the summary statistics in Table~\ref{tab:bgp_combined}. The CDF curves reveal that the ordering of Larger $>$ Exact $>$ Smaller errors holds across the entire error distribution for IP2Location, MaxMind, and IPinfo. DB-IP is an exception where the Larger class produces the lowest errors, consistent with its small Larger prefix pool (n=1,466 observations) noted in Section~\ref{s:prefix:bgp}.

\begin{table}[h!]
\small
\centering
\label{tab:bgp_mobile_fixed}
\resizebox{\columnwidth}{!}{%
\begin{tabular}{llrrrrr}
\toprule
\textbf{Provider} & \textbf{Class} &
\textbf{Mob \%} & \textbf{Mob p50} &
\textbf{Fix \%} & \textbf{Fix p50} &
\textbf{Ratio} \\
\midrule
\multirow{3}{*}{IP2Location}
  & Exact   & 19.4\% & 187~km & 26.0\% &  4~km & 44.0$\times$ \\
  & Smaller & 76.0\% & 246~km & 65.7\% &  3~km & 70.7$\times$ \\
  & Larger  &  4.5\% & 208~km &  8.1\% &  6~km & 36.3$\times$ \\
\midrule
\multirow{3}{*}{MaxMind}
  & Exact   & 13.8\% & 178~km & 27.4\% &  9~km & 18.9$\times$ \\
  & Smaller & 56.8\% & 353~km & 54.0\% & 11~km & 31.1$\times$ \\
  & Larger  & 29.1\% & 217~km & 18.0\% & 28~km &  7.7$\times$ \\
\midrule
\multirow{3}{*}{IPinfo}
  & Exact   &  9.8\% & 111~km & 21.3\% &  6~km & 19.6$\times$ \\
  & Smaller & 78.2\% & 269~km & 68.0\% & 14~km & 19.0$\times$ \\
  & Larger  & 11.9\% & 225~km & 10.5\% &  7~km & 34.5$\times$ \\
\midrule
\multirow{3}{*}{DB-IP}
  & Exact   & 19.4\% & 210~km & 24.6\% & 10~km & 20.7$\times$ \\
  & Smaller & 77.1\% & 262~km & 68.3\% & 26~km & 10.1$\times$ \\
  & Larger  &  3.4\% & 338~km &  6.8\% &  6~km & 52.3$\times$ \\
\bottomrule
\multicolumn{7}{l}{\footnotesize Mob/Fix \%: share of unique prefixes in each
class. Ratio: mobile p50 / fixed p50.} \\
\multicolumn{7}{l}{\footnotesize Mobile: 4,348 obs \quad Fixed: 12,509 obs.
Excludes Both, Satellite, Other ASN types.} \\
\end{tabular}%
}
\caption{Mobile vs.\ fixed median geolocation error by BGP prefix class (IPv4). The mobile/fixed gap persists across all BGP classes and all providers, indicating that BGP prefix coarseness alone does not fully account for the mobile accuracy gap.}
\end{table}

\end{document}